\documentstyle{ioplppt}

\newcommand{\bfA}{{\bf A}}
\newcommand{\bfB}{{\bf B}}
\newcommand{\bfD}{{\bf D}}
\newcommand{\bfG}{\mbox{\boldmath $\Gamma$}}
\newcommand{\bfK}{{\bf K}}
\newcommand{\bfL}{{\bf L}}
\newcommand{\rarr}{\rightarrow}
\newcommand{\sep}{{\;}}
\newcommand{\diag}{\mathop{\rm diag}\nolimits}

\begin{document}

\jl{1}  % J. Phys. A

\bibliographystyle{/users/goliath/lib/tex/mg-ioplike}
  
\title{Lax pair tensors in arbitrary dimensions}[Lax pair tensors]

\author{Martin Goliath\dag\ddag,
  Max Karlovini\dag\S, and
  Kjell Rosquist\dag\P \\[5mm]
  {\dag \ Department of Physics, Stockholm University,}\\
  {\ \  Box 6730, S-113 85 Stockholm, Sweden.}\\
  {\ddag \ E-mail address: goliath@physto.se}\\
  {\S \ E-mail address: max@physto.se}\\
  {\P \ E-mail address: kr@physto.se}
  }

\begin{abstract}
  A recipe is presented for obtaining Lax tensors for any $n$-dimensional
  Hamiltonian system admitting a Lax representation of dimension
  $n$. Our approach is to use the Jacobi geometry and
  coupling-constant metamorphosis to obtain a geometric Lax
  formulation. We also exploit the results to construct integrable
  spacetimes, satisfying the weak energy condition.
\end{abstract}


\maketitle

\section{Introduction}
In this paper we extend the geometric formulation of the Lax pair
equation given in \cite{art:Rosquist1997,art:RosquistGoliath1998}. In
\cite{art:RosquistGoliath1998} a canonical transformation was used to
formulate the 3-particle non-periodic Toda system as geodesic
equations of a 3-dimensional Riemannian space.  However, the canonical
transformation which was used depends on the particular system and so
the method has no obvious extension to more general situations.
A standard way to geometrize a system is to reparametrize the time
variable leading to the Jacobi geometry, see
e.g. \cite{book:Lanczos1986}. This approach has the advantage that it 
always works for Hamiltonians of the type $H=T+V$ with a quadratic 
kinetic energy.  However, it has not been known how to transform the Lax
representation to the Jacobi time gauge.  To remedy this situation we
give a recipe for transforming any Lax representation to the Jacobi
time gauge by using the method of coupling constant metamorphosis 
\cite{art:Hietarinta-et-al1984}.  It turns out that the
resulting Lax system is again homogeneous of degree one in the
momenta.  However, unlike the previous examples it has a non-linear
dependence on the momenta with some terms being proportional to the
square root of the Jacobi Hamiltonian.  As a result the original
geometric formulation of the Lax pair equation \cite{art:Rosquist1997}
cannot be used as it stands in this context. Instead one is lead to a
slightly more general geometric Lax representation.

The approach to geometric formulation of integrable systems presented
in this paper has both advantages and drawbacks. The advantages are
that the dimension of the system is unchanged, that the metric is just
a conformal rescaling of the original kinetic metric (usually flat
space) and that the method works for any integrable system with
quadratic kinetic energy and with an arbitrary number $n$ of
particles. A requirement for the method to work is that the
corresponding Lax representation is of dimension $n$. It is a drawback
that the time must be reparametrized and also that the geometric Lax
formulation involves two dynamical tensors instead of one as was the
case in the original formulation. In any case we consider the present
work as a further step towards a more complete understanding of
integrable geometries and their associated Lax systems. 

We also apply our results to construct two types of integrable
spacetimes satisfying the weak energy condition in an open region.

\section{Transforming Lax pair representations to Jacobi time}
\label{sec:LaxJacobi}
A common feature of completely integrable Hamiltonian systems is the
existence of a Lax pair, i.e. a pair of matrices $(\bfL,\bfA)$
satisfying the equations of motion
\begin{equation}\label{eq:laxeq}
  \frac{d\bfL}{dt}=[\bfL,\bfA] ,
\end{equation}
where the time derivative is defined by
\begin{equation}
  \frac{d\bfL}{dt} = \{\bfL,H\} .
\end{equation}
It follows that $I_k=k^{-1}\Tr\bfL^k$, $k=1,\,2,\ldots$, is a sequence
of invariants of the system. We consider $n$-dimensional Hamiltonians
of the classical type 
\begin{equation}\label{eq:Hclassical}
  H=T+V , \quad
  T=\textstyle\frac12h^{\alpha\beta}p_\alpha p_\beta , \quad
  V=V(q) , 
\end{equation}
where $h_{\alpha\beta}$ is the kinetic metric. Furthermore, we
restrict to Lax matrices of dimension $n$  that are linear (but not
necessarily homogeneous) in the momenta: 
\begin{equation}\label{eq:laxmatrices}
  \begin{array}{r@{\sep}l}
    \bfL & = \bfL^\alpha p_\alpha + \bfK , \\[3mm]
    \bfA & = \bfA^\alpha p_\alpha + \bfD .
  \end{array}
\end{equation}
Here, $\bfL^\alpha$, $\bfA^\alpha$, $\bfK$ and $\bfD$ are independent
of the momenta. We seek a general recipe for transforming such a Lax
representation under the Jacobi time transformation $t\rarr t_J$,
$dt_J=2(E-V)dt$, which maps orbits of an energy surface $H=E$ into
geodesics of the Jacobi geometry
$g_{\alpha\beta}=2(E-V)h_{\alpha\beta}$, i.e. into orbits of the
Jacobi Hamiltonian
$H_J=\frac12 g^{\alpha\beta}p_\alpha p_\beta=[2(E-V)]^{-1}T$. This is
naturally accomplished by performing the time transformation as a
coupling constant metamorphosis \cite{art:Hietarinta-et-al1984}. To
this end we first of all need to introduce a coupling constant into
our original Hamiltonian. This is accomplished by making the
rescalings $p_\alpha \rarr \lambda^{-1} p_\alpha$ and $H \rarr
\lambda^{-2} H$, resulting in
\begin{equation}
  H = T + \lambda^2 V,
\end{equation}
where $T$ is defined by its original functional form. The
transformation is noncanonical, yet when accompanied by the time
rescaling $t\rarr \lambda t$, the canonical equations of motion are
preserved. However, the same does not hold true for the Lax equation
(\ref{eq:laxeq}), but this is a drawback which just as easily can be
cured by making another rescaling, namely
$\bfA\rarr\lambda^{-1}\bfA$. Finally, we fix the gauge freedom to make 
arbitrary rescalings of $\bfL$ by letting
$\bfL\rarr\lambda^{-1}\bfL$, which gives the rescaled Lax pair 
\begin{equation}\label{eq:laxmatrices2}
  \begin{array}{r@{\sep}l}
    \bfL & = \bfL^\alpha p_\alpha + \lambda\bfK , \\[3mm]
    \bfA & = \bfA^\alpha p_\alpha + \lambda\bfD .
  \end{array}
\end{equation}
The passage to the Jacobi time $t_J$ can now be obtained as a coupling
constant metamorphosis acting on the coupling constant
$\kappa:=\frac12\lambda^2$ which has entered the Hamiltonian. However, to
end up with the homogeneous Hamiltonian $H_J$ when solving a fixed
energy constraint for $\kappa$, we need to hold on to the
interpretation of the parameter $E$ as the energy value of the
original Hamiltonian $H=T+V$, thereby making $\lambda^2E$ the
corresponding energy value of the rescaled Hamiltonian
$H=T+\lambda^2V$. This in fact means that we are not really dealing
with a coupling constant metamorphosis in the original sense, as the
old energy $E$ does not enter linearly into the new Hamiltonian
$H_J$. Nevertheless, it is not difficult to realise that the results
of \cite{art:Hietarinta-et-al1984} still apply. It thus follows that
when substituting $H_J$ for $\kappa$ (i.e. $\sqrt{2H_J}$ for
$\lambda$) in the expressions for $H$, $\bfL$ and $\bfA$ above, the
original Lax equation (\ref{eq:laxeq}) can be written as 
\begin{equation}
  \frac{d\bfL}{dt_J}=[2(E-V)]^{-1}[\bfL,\bfA] ,
\end{equation}
where
\begin{equation}
  \frac{d\bfL}{dt_J}=\{\bfL,H_J\} ,
\end{equation}
and the factor $[2(E-V)]^{-1}$ in front of the matrix commutator arises
from the fact that $d/dt=2(E-V)d/dt_J$. Absorbing this factor into
$\bfA$, we obtain the final time transformed Lax equation. To
summarise, the original Lax pair system with $H$, $\bfL$ and
$\bfA$ given by equation (\ref{eq:Hclassical}) and equation
(\ref{eq:laxmatrices}), is transformed to the Jacobi time gauge
according to 
\begin{equation}
  \begin{array}{r@{\sep}l}
    H_J&=[2(E-V)]^{-1}T , \\[3mm]
    {\bfL}_J&=\bfL^\alpha p_\alpha+\sqrt{2H_J}\,\bfK , \\[3mm]
    {\bfA}_J&=[2(E-V)]^{-1}(\bfA^\alpha p_\alpha+\sqrt{2H_J}\,\bfD) .
  \end{array}
\end{equation}
Note that all of these objects are homogeneous in the momenta although
the two Lax matrices are not polynomials. In the following 
we will suppress the index $J$ referring to the Jacobi time and absorb 
the factor $[2(E-V)]^{-1}$ into the definitions of $\bfA^\alpha$ and
$\bfD$, thus focusing on the geodesic Lax pair systems given by
\begin{equation}\label{eq:JLax}
  \begin{array}{r@{\sep}l}
    H    & =\textstyle\frac12 g^{\alpha\beta}p_\alpha p_\beta , \\[3mm]
    \bfL & =\bfL^\alpha p_\alpha+\sqrt{2H}\,\bfK , \\[3mm]
    \bfA & =\bfA^\alpha p_\alpha+\sqrt{2H}\,\bfD .
  \end{array}
\end{equation}

\section{The Lax pair tensors}
In \cite{art:Rosquist1997}, a geometrical formulation of the Lax equation was
found by demanding that the tensors corresponding to the Lax matrices should
be linear and also homogeneous in the momenta.  We now generalise this
geometrization by taking the system (\ref{eq:JLax}) as our starting point.
This leads us to write the components of the Lax matrices with mixed indices
({\it i.e.} row indices contravariant and column indices covariant) so that
$\bfL = (L^\alpha{}_\beta)$, $\bfA = (A^\alpha{}_\beta)$, $\bfK =
(K^\alpha{}_\beta)$ and $\bfD = (K^\alpha{}_\beta)$.  We then extract the
geometrical objects $L^\alpha{}_\beta{}^\gamma$, $A^\alpha{}_\beta{}^\gamma$,
from the components of (\ref{eq:JLax})
\begin{equation}\label{LKAD}
  \begin{array}{r@{\sep}l}
    L^\alpha\!_\beta & =L^\alpha\!_\beta\!^\gamma p_\gamma+
    \sqrt{2H}\,K^\alpha\!_\beta \ , \\[3mm]
    A^\alpha\!_\beta & =A^\alpha\!_\beta\!^\gamma p_\gamma+
    \sqrt{2H}\,D^\alpha\!_\beta \ .
  \end{array}
\end{equation}
Next we define connection matrices by $\bfG^\gamma =
(\Gamma^\alpha{}_\beta{}^\gamma)$ where $\Gamma^\alpha{}_\beta{}^\gamma =
g^{\gamma\delta} \Gamma^\alpha{}_{\beta\delta}$ and
$\Gamma^\alpha{}_{\beta\gamma}$ is the Levi-Civita connection associated to
the Jacobi metric.  In the following, the Jacobi metric will be used to raise
and lower indices in this way on other objects as well.  Judging from the
original Lax tensor formulation we expect that $\bfL^\alpha$ and thereby also
$\bfK$ will be tensors.  Anticipating this result we replace partial
derivatives with covariant derivatives according to recipes
\begin{equation}\label{eq:derivatives}
  \begin{array}{r@{\sep}l}
    \bfL_{\alpha,\beta} & =\bfL_{\alpha;\beta}+
    \left[\bfL_\alpha,\bfG_\beta\right]-
    \Gamma_\alpha\!^\gamma\!_\beta\bfL_\gamma , \\[3mm]
    \bfK_{,\alpha} & =\bfK_{;\alpha}+\left[\bfK,\bfG_\alpha\right] .
  \end{array}
\end{equation}
This procedure will be justified below.  Using (\ref{eq:laxmatrices}), the
commutator $[\bfL,\bfA]$ of the Lax equation can be written in the form
\begin{equation}
    \left(\left[\bfL^\alpha,\bfA^\beta\right]+
    \left[\bfK,\bfD\right]g^{\alpha\beta}\right)p_\alpha p_\beta
     +\sqrt{2H}\,p_\alpha\left(\left[\bfL^\alpha,\bfD\right]+
    \left[\bfK,\bfA^\alpha\right]\right) . \label{eq:com}
\end{equation}
The Poisson bracket $\{\bfL,H\}$, on the other hand, using
(\ref{eq:derivatives}), becomes
\begin{equation}
  \bigl(\bfL^{\alpha;\beta}+\bigl[\bfL^\alpha,\bfG^\beta\bigr]\bigr)
  p_\alpha p_\beta
  +\sqrt{2H}\,p_\alpha\left(\bfK^{;\alpha}
  +\left[\bfK,\bfG^\alpha\right]\right) . \label{eq:poi}
\end{equation}
Identifying terms in (\ref{eq:com}) and (\ref{eq:poi}) yields the
following form of the Lax equations:
\begin{equation}\label{eq:laxeq2}
  \begin{array}{r@{\sep}l}
    \bfL_{(\alpha;\beta)} & =
    \left[\bfL_{(\alpha},\bfB_{\beta)}\right]+
    \left[\bfK,\bfD\right]g_{\alpha\beta} ,\\[3mm]
    \bfK_{;\alpha} & =
    \left[\bfL_\alpha,\bfD\right]+
    \left[\bfK,\bfB_\alpha\right] ,
  \end{array}
\end{equation}
where we have defined $\bfB^\alpha = \bfA^\alpha - \bfG^\alpha$.  This shows
that we can consistently interpret $\bfL$, $\bfB$, $\bfK$ and $\bfD$ as
tensors since equations (\ref{eq:laxeq2}) are then manifestly covariant.
However, note that $\bfA^\alpha$ is a connection-like object.  By setting
$\bfK=0$, $\bfD=0$, the second of equations (\ref{eq:laxeq2}) becomes an
identity while the first equation reduces to the Lax tensor equation of
\cite{art:Rosquist1997}.

\section{Examples}
We now apply the above geometrized Lax formulation to some systems
with known Lax representations. They are all of the form
(\ref{eq:Hclassical}) with flat kinetic metric
$h_{\alpha\beta}=\delta_{\alpha\beta}$. For the systems considered
below, it is found that the matrix $\bfL_*:=\bfL^\alpha p_\alpha$ is
diagonal, while $\bfK$ has no diagonal elements. It follows that
$\Tr\bfK=0$ and $\Tr(\bfL_*\bfK)=0$. In addition, $\bfL_*$ and $\bfK$
satisfy  
\begin{equation}
  \Tr\bfL_*=\sum_\alpha p_\alpha , \quad
  \Tr(\bfL_*^2)=2T , \quad
  \Tr(\bfK^2)=2V , 
\end{equation}
so that the first two invariants of the geometrized system are
\begin{equation}
  \begin{array}{r@{\sep}l}
    I_1 & =\Tr\bfL=\sum_\alpha p_\alpha , \\[3mm]
    I_2 & =\frac{1}{2}\Tr(\bfL^2)=T+2HV=\frac{ET}{E-V}=2EH .
  \end{array}
\end{equation}
Also, the metric for these models is given by
\begin{equation}
  g^{\alpha\beta}=\frac{\Tr(\bfL^\alpha\bfL^\beta)}{2E-\Tr(\bfK^2)} ,
\end{equation}
where $\Tr(\bfL^\alpha\bfL^\beta)=h^{\alpha\beta}$.

The results obtained below for the different models have been verified
for the case $n=3$, using the package GRTensorII \cite{man:GRTensor1996}
for MapleV.

\subsection{The Toda lattice}
First, we will consider the $n$-particle non-periodic (open) Toda
lattice, for which  
\begin{equation}\label{eq:Toda}
  V=\sum_{i=1}^{n-1} a_i\!^2 , \quad
  a_i=\exp(q^{i}-q^{i+1}) . 
\end{equation}
A standard Lax representation of this system is \cite{book:Perelomov1990} 
\begin{equation}
  \begin{array}{r@{\sep}l}
    \bfL=L^i\!_j= &
    \sum_{k=1}^n p_k\delta^i\!_k    \delta_{j\,k}+
           a_k\left(\delta^i\!_k    \delta_{j,k+1}+
                    \delta^i\!_{k+1}\delta_{j\,k}\right) ,\\[3mm]
    \bfA=A^i\!_j= &  \sum_{k=1}^n
           a_k\left(\delta^i\!_k    \delta_{j,k+1}-
                    \delta^i\!_{k+1}\delta_{j\,k}\right) .
  \end{array} 
\end{equation}
From this Lax representation we find
\begin{equation}
  \begin{array}{l}
    \bfL^\alpha=L^i\!_j\!^\alpha=
    \delta^{i\,\alpha}\delta_j\!^\alpha , \\[3mm]
    \bfK=K^i\!_j=\sum_{k=1}^n 
    a_k\left(\delta^i\!_k    \delta_{j,k+1}+
             \delta^i\!_{k+1}\delta_{j\,k}\right) , \\[3mm]
    {
      \bfA^\alpha=0 , \quad
      \bfB^\alpha=-\bfG^\alpha , \quad
      \bfD=\frac{1}{2(E-V)}\bfA } .
  \end{array}
\end{equation}
Note that the factor $[2(E-V)]^{-1}$ is absorbed into $\bfD$, as
discussed in section \ref{sec:LaxJacobi}.

We note in passing that the results for a periodic Toda lattice can be
obtained by letting the sum in (\ref{eq:Toda}) run from 1 to $n$ and
by employing the cyclicity conditions 
\begin{equation}\label{eq:cyc}
  q^{n+1}\rarr q^1 , \quad
  \delta^i\!_{n+1}\rarr \delta^i\!_1 , \quad
  \delta_{j,n+1}\rarr \delta_{j\,1} .
\end{equation}

\subsection{The Calogero-Moser system}
Our second example is the Calogero-Moser system (the type I system of
\cite{art:OlshanetskyPerelomov1981}). The potential is 
\begin{equation}
  V=\sum_{i<j} a_{i\,j}^{\;2} , \quad
  a_{i\,j}=\frac{1}{q^i-q^j} . 
\end{equation}
A well-known Lax representation of this system is
\cite{art:OlshanetskyPerelomov1981} 
\begin{equation}
  \begin{array}{r@{\sep}l}
    \bfL=L^i\!_j= &
    \sum_{k=1}^n p_k\delta^i\!_k\delta_{j\,k} +
    i(1-\delta^i\!_j)a_{i\,j} , \\[3mm]
    \bfA=A^i\!_j= &
    i\left(\delta^i\!_j\sum_{k\neq i}a_{i\,k}^{\;2}-
    (1-\delta^i\!_j)a_{i\,j}^{\;2}\right) , \\[3mm]
  \end{array}
\end{equation}
from which it follows that 
\begin{equation}
  \begin{array}{l}
    \bfL^\alpha=L^i\!_j\!^\alpha=
    \delta^{i\,\alpha}\delta_j\!^\alpha , \\[3mm]
    \bfK=K^i\!_j=i(1-\delta^i\!_j)a_{i\,j} , \\[3mm]
     {
      \bfA^\alpha=0 , \quad
      \bfB^\alpha=-\bfG^\alpha , \quad
      \bfD=\frac{1}{2(E-V)}\bfA } .
  \end{array}
\end{equation}

\section{Four-dimensional spacetime generalisations}
In general relativity, symmetries of spacetime play an important
role. Many examples of spacetimes with linear invariants,
corresponding to Killing vectors, are known
\cite{book:ExactSolutions1980}. However , higher-order invariants
(apart from the trivial quadratic invariant
$g^{\alpha\beta}p_\alpha p_\beta$, associated with the metric) are
quite rare. A well-known example is the second-rank Killing tensor of
the Kerr spacetime \cite{art:WalkerPenrose1970}, which together with
the two existing Killing vectors enable a complete integration of the
geodesics of that system. Examples of spacetimes with a non-trivial
third-rank Killing tensor were given in
\cite{art:RosquistGoliath1998}, but apart from that, such Killing
tensors are to our knowledge unknown. In this perspective, it is of
interest to extend the Riemannian geometries obtained above (with
$n=3$) to four-dimensional spacetimes that inherit the 
symmetries of the original geometry. The simplest generalisation is
obtained by introducing a timelike coordinate $q^0$ according to
\begin{equation}\label{eq:ds4}
  ^{(4)}ds^2 = -(dq^0)^2 + ds^2 ,
\end{equation}
where $ds^2=2(E-V)h_{\alpha\beta}dq^\alpha dq^\beta$,
$\alpha , \beta \in \{1,2,3\}$, and $V$ is one
of the potentials studied above. Note that the condition $E>V$
must be satisfied for the metric to have Lorentzian signature. This is 
not a serious drawback, since the sign of $E-V$ is preserved along the 
geodesics of the metric (\ref{eq:ds4}). Hence a completely
integrable (1+3)-dimensional geometry is well-defined in the region
$E>V$. In general, these spacetimes are of 
Petrov type II. Choosing a particular potential $V$, the Petrov type
may be further specialised. The Toda and Calogero-Moser potentials
both lead to Petrov type II spacetimes. In a Lorentzian frame
corresponding to the given coordinates, the energy-momentum tensor
$T^{ab}$, $a,b\in\{0,1,2,3\}$, for a spacetime of this type satisfies
$T^{0\alpha}=0$ and can thus be diagonalised using a rotation of the
spatial part of the frame. Another characteristic feature is the fact that
the eigenvalue equation for the spatial part of $T^{ab}$ naturally
factorises into one linear and one second-degree equation. This is due 
to the existence of the space-like Killing vector
$\Tr\bfL^\alpha$. The linear equation gives the anisotropic pressure
in the direction of this Killing vector. In terms of
$\tilde{T}^{ab}=2(E-V)^3T^{ab}$, the Lorentzian frame
components of the energy-momentum tensor for the Toda spacetime become
\begin{equation}
  \begin{array}{r@{\sep}l}
    \tilde{T}^{00} & =
    8EV-2\sum_{i=1}^3a_i\!^2(a_i\!^2+11a_{i+1}^{\;2}), \\[3mm]   
    \tilde{T}^{\alpha\alpha} & =
    -2(V+a_{\alpha+1}^{\;2})E+(V+9a_{\alpha+1}^{\;2})(V-a_{\alpha+1}^{\;2}),
    \\[3mm]
    \tilde{T}^{\alpha,\alpha+1} & =
    (-2E+5V-6a_\alpha\!^2)a_\alpha\!^2-3a_{\alpha-1}^{\;2}a_{\alpha+1}^{\;2} ,
  \end{array}
\end{equation}
where it is to be understood that indices are added modulo $3$. These
expressions hold for the open case ($a_3=0$) as well as for the
closed case ($a_1a_2a_3=1$). Diagonalising the energy-momentum tensor gives
$\tilde{T}^{ab}=\diag{(\tilde{\mu},\tilde{p_1},\tilde{p_2},\tilde{p_3})}$,
where
\begin{equation}
  \begin{array}{r@{\sep}l}
    \tilde{\mu} = & \tilde{T}^{00}, \\[3mm]
    \tilde{p}_1 = & -4EV + 12\sum_{i=1}^3 a_i\!^2a_{i+1}^{\;2}, \\[3mm] 
    \tilde{p}_{2,3} = & \Lambda \pm \sqrt{\Lambda^2-\Delta}, \\[3mm]
    \Lambda = & -2EV + \sum_{i=1}^3a_i\!^2(a_i\!^2 +  5a_{i+1}^{\;2}), \\[3mm]
    \Delta = & 12(-13E+9V)a_1\!^2a_2\!^2a_3\!^2 
    +2\sum_{i=1}^3a_i\!^2\{6E^2a_{i+1}^{\;2} - \\[3mm]
    & E[4a_i\!^4 + 3a_{i+1}^{\;2}(a_i\!^2 + a_{i+1}^{\;2})] +              
    9a_{i+1}^{\;2}(a_i\!^2 - a_{i+1}^{\;2})^2\} .
  \end{array}
\end{equation}
Since the eigenvalues of $T^{ab}$ are related to those of
$\tilde{T}^{ab}$ by a positive factor whenever the metric signature
condition $E>V$ holds, the weak energy condition
\cite{book:HawkingEllis1973} reads $\tilde{\mu}\geq 0$,
$\tilde{\mu}+\tilde{p}_\alpha\geq 0$ which is equivalent to
$\tilde{\mu}\geq 0$, $\tilde{\mu}+\tilde{p}_1\geq 0$,
$\tilde{\mu}+\Lambda\geq 0$, $\Delta\geq 0$. At the spatial origin
$q^\alpha=0$, where $V=2$ ($V=3$) for the open (closed) case, these
inequalities are  
satisfied if $E>2$ ($E>3$). Hence by continuity, there
must be some open region in $(E,q^\alpha)$-space where the metric
signature condition and the weak energy condition hold simultaneously. 

Calculating $T^{ab}$ for the Calogero-Moser spacetime yields the result 
\begin{equation}
  \begin{array}{r@{\sep}l}
    \tilde{T}^{00} = & 3\sum_{i=1}^3 4Ea_i\!^4 -
    (a_i\!^3+a_{i+1}^{\;3})^2 
    -4a_i\!^2a_{i+1}^{\;2}(a_i\!^2+a_{i+1}^{\;2}), \\[3mm]
    \tilde{T}^{\alpha\alpha} = &
    -3E(a_{\alpha-1}^{\;4}+a_\alpha\!^4+2a_{\alpha+1}^{\;4})
    +3(a_{\alpha-1}^{\;4}+a_\alpha\!^4)a_{\alpha+1}^{\;2} \\[3mm]
    &+3(V-a_{\alpha+1}^{\;2})(a_{\alpha-1}^{\;2}a_\alpha\!^2+
    2a_{\alpha+1}^{\;4})-2a_{\alpha-1}^{\;3}a_\alpha\!^3 \\[3mm]
    &+4(a_{\alpha-1}^{\;3}+a_\alpha\!^3)a_{\alpha+1}^{\;3} 
    + 2\sum_{i=1}^3a_i\!^6, \\[3mm]
    \tilde{T}^{\alpha,\alpha+1} = & -3a_\alpha\!^4(E-V+a_\alpha\!^2)
    +3a_\alpha\!^3(a_{\alpha-1}^{\;3}+a_{\alpha+1}^{\;3})\\[3mm]
    &-3a_{\alpha-1}^{\;3}a_{\alpha+1}^{\;3},
  \end{array}
\end{equation}
where $\tilde{T}^{ab}$ is defined as in the Toda case. 
As a Calogero-Moser system with three particles can be viewed as a
system with nearest-neighbour interaction, we have used the notation
$a_i=(q^i-q^{i+1})^{-1}$, $V=\sum_{i=1}^3 a_i{}^2$. The eigenvalues of  
$\tilde{T}^{ab}$ can be written down in a form analogous to the Toda
case. We choose not to do so here, however, as the expressions are
more complicated and not very illustrative. As for the Toda system,
there is a region where the weak energy condition is satisfied. 
This can be verified, e.g. by setting $q^1=-1$, $q^2=0$, $q^3=1$,
which gives $V=9/4$ and the following eigenvalues for
$\tilde{T}^{ab}$: 
\begin{equation}
  \begin{array}{r@{\sep}l}
    \tilde{\mu} = & \frac{99}{4}(E-\frac{545}{264}), \\[3mm]
    \tilde{p}_1 = & -\frac{99}{8}(E-\frac{793}{396}), \\[3mm] 
    \tilde{p}_{2,3} = & -\frac{99}{16}(E-\frac{421}{198})\pm
    \frac{3}{16}|15E-46| . \\[3mm] 
  \end{array}
\end{equation}
For $E>9/4$, these eigenvalues satisfy the weak energy condition.

\section{Discussion}
In this paper we have presented a general procedure for obtaining a
tensorial Lax representation from known Lax pair matrices. The
tensorial representation is obtained via a time reparametrization. It
has two dynamical Lax tensors ($\bfL$ and $\bfK$), instead of only
$\bfL$ as in the original geometric formulation
\cite{art:Rosquist1997}. One advantage with this particular
geometrical formulation is that it provides a straight-forward recipe
for obtaining Lax tensors from a known Lax representation. Indeed, any
$n$-dimensional Hamiltonian system of the classical form
(\ref{eq:Hclassical}) that has a Lax representation of dimension $n$
can be geometrized in this way.

\end{document}